\documentclass[preprint2]{aastex}
\usepackage[usenames]{color}

\slugcomment{}

\shorttitle{The chromospherically active binaries V789~Mon and GZ~Leo
}
\shortauthors{G\'alvez et al.}

\begin{document}


\title{Multiwavelength optical observations of two chromospherically
active binary systems: V789~Mon and GZ~Leo\altaffilmark{*}}


\author{M.C. G\'alvez\altaffilmark{1, 2}, 
D.~Montes\altaffilmark{2}, 
M.J.~Fern\'{a}ndez-Figueroa\altaffilmark{2}, 
E.~De Castro\altaffilmark{2} 
and M.~Cornide\altaffilmark{2}}


\altaffiltext{1}{Centre for Astrophysics Research, Science and Technology Research Institute, 
University of Hertfordshire, Hatfield AL10 9AB, UK}
\altaffiltext{2}{Departamento de Astrof\'{\i}sica,
Facultad de Ciencias F\'{\i}sicas,
 Universidad Complutense de Madrid, E-28040 Madrid, Spain}
\email{mcz@astrax.fis.ucm.es}
\altaffiltext{*}{Based on observations collected with the 2.2~m telescope
at the Centro Astron\'omico Hispano Alem\'an (CAHA) at Calar
 Alto (Almer\'{\i}a, Spain), operated jointly
by the Max-Planck Institut f\"{u}r Astronomie and the Instituto de
Astrof\'{\i}sica de Andaluc\'{\i}a (CSIC) and
 with the 2.1~m Otto Struve Telescope at McDonald Observatory of
 the University of Texas at Austin (USA)}



\begin{abstract}
 This paper describes a multiwavelengh optical study
of chromospheres in two X-ray/EUV selected
active binary stars with strong H$\alpha$ emission,
 \object{V789~Mon (2RE J0725-002)} and \object{GZ~Leo (2RE J1101+223)}.
The goal of the study is to determine radial velocities and
fundamental stellar parameters in chromospherically active binary
systems in order to include them in the activity-rotation
and activity-age relations.
We carried out high resolution echelle spectroscopic observations and applied
spectral subtraction technique in order
to measure emission excesses due to chromosphere.
The detailed study of activity indicators allowed us to characterize
the presence of different chromospheric features in these systems
and enabled to include them in a larger activity-rotation survey.
We computed radial velocities of the systems using cross
correlation with the radial velocity standards.
The double-line spectral binarity was confirmed and
the orbital solutions improved for both systems.
In addition, other stellar parameters such as:
spectral types, projected rotational velocities ($v\sin{i}$), and
the equivalent width of the lithium Li~{\sc i} $\lambda$6707.8 \AA \
 absorption line were determined.
\end{abstract}



\section{Introduction}

Why each particular star reaches a certain level
of activity, and why seemingly similar stars
may have drastically different chromospheres?
It is believed that differential rotation and convection
are the key factors responsible for the magnetic activity in the star.
A significant fraction of stellar population, especially
at later evolutionary stages, is found in close binaries,
where tidal forces and mass transfer become
additional factors controlling rotation
and atmospheric properties of companions.

In order to understand the physical origin of activity,
the time-scale of variability,
and to be able to extrapolate to other systems,
we initiated a multiwavelengh optical survey
of a large sample of binaries with various levels of activity.
We carried out the study of the chromosphere of active binary systems
 using the information provided by several optical and near infrared spectroscopic
  features that are formed at different heights in the chromosphere 
 (see Montes et al. 1997, 1998, 2000; G\'alvez et al. 2002, 2007),
and applied it to several known and newly discovered binaries,
most of which are BY Dra or RS CVn type
with various levels of activity.

The ultimate goal of the project is to establish
a reliable activity-rotation relation in binary stars.
To achieve high accuracy one needs to investigate all
possible causes of activity for each particular system
and for each component in the system.
It is essential to take into consideration
dispersion and signal-to-noise of the spectral data
obtained with different instrumental set-ups.
Furthermore, the study of binaries as compared to single stars
is complicated by the interaction between companions,
line blending (SB2 case), reflected light,
etc. Active stars, which are recognized by the presence
of emission lines in one or both components,
present the most challenging case
and need to be studied on the individual basis.
To obtain an accurate orbital solution and
to study the behavior of activity,
one needs to separate radial velocities, 
 equivalent widths ($EW$ hereafter) of emission lines,
and fluxes between two components. It is also vital
to carry out a precise stellar classification due to the importance of
the subgiant and giant presence in the activity-rotation relation
(e.g. in RS CVn case).

Here we concentrate on two recently discovered, X-ray/EUV selected, chromospherically active
 double-line spectroscopic binaries included in this survey:
 V789~Mon (2RE J0725-002, BD-00 1712) and GZ~Leo (2RE J1101+223,
HD~95559, BD+23 2297, HIP 53923).

V789~Mon, $V=9.33$, was classified by Jeffries et al.
(1995)  as a  double-line spectroscopic binary (SB2) with  almost
identical K5V components. They gave a rotational velocity, $v\sin{i}$,
of 25 km\,s$^{-1}$ and an orbital period of 1.40 days.
 The photometric period (1.412 days) reported by  Cutispoto et al.
(1999) and Robb \& Gladders (1996) indicates synchronous rotation.
Jeffries et al. (1995) suggest that this system could be eclipsing,
but Robb \& Gladders (1996) did not find any evidence of eclipses in
their photometric observations.
Cutispoto et al. (1999) derived a K3V + M0:V classification for this
system from colors and spectral signatures. Regarding activity,
Jeffries et al. (1995) detected the H$\alpha$ line in emission above the
continuum level for both components.

GZ~Leo is a SB2 binary with $V=8.8$ mag.
 Jeffries et al. (1995) obtained a first orbital solution with
a period of 1.528 days using eleven radial velocity measurements. Later,
Fekel \& Henry (2000) reported an improved orbital solution using nine
new radial velocity determinations and one additional value given
by Strassmeier et al. (2000).
The system consists of two similar rapidly rotating stars
($v\sin{i}$ $\approx$ 30 km\,s$^{-1}$). The spectral type of
the components is not well established. Jeffries et al. (1995) classified it
as G5 while Popper (1996) as K2, based on the strength
of the Na~{\sc i} D$_{1}$ \& D$_{2}$ lines.
The latter is closer to the K1V type determined
by Fekel \& Henry (2000) from the temperature and
luminosity-sensitive line ratios in the 6430-6465 \AA\ region.
Popper (1996) reported photometric variability, but could not
confirm whether it is due to eclipses.
Strassmeier et al. (2000) determined a period of 2.944 days from their
photometric observations, but noted that a period of 1.514 days was also
possible. Finally, Fekel \& Henry (2000) and Pandey et al. (2002, 2005)
confirmed that the photometric period (1.526 days) is nearly identical to
its orbital period.
This binary system shows the typical signatures of chromospheric
activity. A filled-in H$\alpha$ absorption line in both components
was reported by Jeffries et al. (1995), while Ca~{\sc ii} H \& K
emissions were detected in the spectra by Strassmeier et al. (2000).
In addition, GZ~Leo is a young system as indicated by its large
lithium abundance (Strassmeier et al. 2000; Fekel \& Henry 2000).

A preliminary analysis of these two systems can be found in G\'alvez et
al. (2003) and G\'alvez (2005).

In $\S$~2 we summarize our observations.
In $\S$~3 we derive stellar parameters and the
orbit elements of the systems.
The behavior of the different chromospheric
activity indicators is described in $\S$~4.
Finally $\S$~5 gives a summary.

\section{Observations}

In this paper we analyzed high resolution echelle spectra of both
systems that we obtained during two observing runs detailed
in Table~\ref{tab:reduc1}.
The spectra were extracted using the standard
reduction procedures in the
IRAF\footnote{IRAF is distributed by the National Optical Observatory,
which is operated by the Association of Universities for Research in
Astronomy, Inc., under contract with the National Science Foundation.}
 echelle package (bias subtraction, flat-field division and 
 optimal extraction
 of the spectra). We obtained the wavelength calibration by taking
 spectra of a Th-Ar lamp. Finally, we normalized the spectra by
 a polynomial fit to the observed continuum.

We obtained nine spectra of V789~Mon in one observing run and fifteen
 spectra of GZ~Leo in two. The signal to noise ratio ($S/N$) achieved
 in the H$\alpha$ line region in all the spectra is $\approx$ 60 for V789~Mon
 and $\approx$ 100 for GZ~Leo.

\section{Stellar parameters of the binary systems}

We list the adopted stellar  parameters of both systems in
Table~\ref{tab:par}. Spectral type and the photometric data
($B-V$, $V$, $P_{\rm phot}$) are taken from SIMBAD and Cutispoto et al.
(1999) for V789~Mon and from SIMBAD and Fekel \& Henry (2000) for
GZ~Leo. The astrometric data (parallax, $\pi$; proper motions,
$\mu$$_{\alpha}$$\cos{\delta}$ and  $\mu$$_{\delta}$) are from $Hipparcos$
(ESA 1997) and $Tycho-2$ (H$\o$g et al. 2000) catalogues.
 For V789~Mon no parallax value is available, so we give the spectroscopic
 parallax determined here (see $\S$~3.4). Finally,
 the orbital period ($P_{\rm orb}$) and the projected rotational velocities
($v\sin{i}$) have been determined in this paper as well (see below).

\subsection{Spectral types}
 In order to obtain the spectral type of these binary systems we
 compared our high resolution echelle spectra
 with spectra of inactive reference stars of different spectral
 types and luminosity classes observed during the same observing
 run.
 To avoid contamination by chromospheric emission,
 we only used spectral orders that are free of lines sensitive
 to chromospheric activity.
This analysis makes use of the program {\sc starmod} developed
at Penn State University (Barden 1985) and modified by us.
{\sc starmod} allows to rotationally broaden spectra and shift
them in the velocity space.
We ran it on the spectra of the reference stars and combined
with the appropriate weights to create
composite spectra that were used as template for comparison with
our targets.

For V789~Mon we obtained the best fit between observed
and synthetic spectra using a K5V reference star (\object{HR~1614})
for both components, contributing respectively 60\% and 40\%
of the total flux in the continuum.  The fits were carried out in
all the phases where the spectra of the components could be separated.
 Our spectral classification is in agreement with a K5V
 type reported by Jeffries et al. (1995), while the relative
 contribution of the companions is compatible with
 Cutispoto et al. (1999) result that assigned a later
 spectral type for the secondary component.

In the case of GZ~Leo, we obtained the best fit between observed and
synthetic spectra using a K0V reference star (\object{HR~166}) for both
 components, with the equal contribution from each component.
 Similarly to V789~Mon, the fits were performed in all the phases where
 the components could be separated, and several spectral types and 
 luminosity class reference stars were used.
 Our spectral classification is very close to the K2V
type reported by Popper (1996) and the K1V type by Fekel \&
Henry (2000).

\subsection{Rotational velocity}
Jeffries et al. (1995) estimated the projected rotational velocity
 ($v\sin{i}$) for V789~Mon as 25\,km\,s$^{-1}$. For
 GZ~Leo they found 27 km\,s$^{-1}$, while Fekel \& Henry (2000)
 reported values of 32.4 and 31.6 km\,s$^{-1}$ for each
 component. Fekel \& Henry (2000)  noted that the actual $v\sin{i}$
 may slightly smaller, because spots cause the lines to be shallower
 than in an unspotted star, making FWHM appear larger.
 Strassmeier et al. (2000) report
 similar values of 31 and 26 km\,s$^{-1}$ respectively.

 Using {\sc starmod} we obtained
for each observing run $v\sin{i}$ values about 30 km\,s$^{-1}$ 
for both components of V789~Mon and GZ~Leo.
 In order to determine a more accurate rotational velocities
we applied a cross-correlation technique to our high
 resolution echelle spectra of these stars using IRAF task {\sc fxcor}.
The method is described in detail in our previous papers (see G\'alvez et al.
 2002; L\'opez-Santiago et al. 2003).
 In short, the method is based on the fact that when a stellar
spectrum with rotationally broadened lines is cross-correlated
against a narrow-line spectrum, the width of the
cross-correlation function (CCF) will depend on the amount of
rotational broadening of the former spectrum.

For V789~Mon, we used as a template a slowly rotating K5V star
HR 1614. We obtained a $v\sin{i}$ of 28.28$\pm$1.59 and 
 25.09$\pm$2.06 km\,s$^{-1}$
for the primary and the secondary component respectively.

In the case of GZ~Leo, we used a K0V star HR 166 and
 a K2V star \object{HD 166620} as templates, in
 McDonald and FOCES observing runs respectively.
 We obtained averaged values of 26.23$\pm$1.13 and 26.92$\pm$1.14 km\,s$^{-1}$
 for each component (see Table~\ref{tab:par}).

\subsection{Radial velocities and orbital solution}
Heliocentric radial velocities were also obtained
 by using the cross-correlation technique
 (see e.g. G\'alvez et al. 2007). The spectra of the
 target were cross-correlated order by order, using the routine
{\sc fxcor} in IRAF, against spectra of radial velocity standards with
similar spectral type taken from Beavers et al. (1979).
We derived the radial velocity for each order from the position of peak of
the
cross-correlation function (CCF) and calculated the uncertainties
 based on the fitted peak height and the antisymmetric noise as
described by Tonry \& Davis (1979).

 Since both systems are SB2, we could see two peaks in the CCF
 associated with two components; we fitted each peak separately.
 When the component peaks were too close, we used de-blending technique.

In the Tables~\ref{tab:bvr} and~\ref{tab:vr} we list, for each spectrum, the
heliocentric radial velocities ($V_{\rm hel}$)
and their associated errors ($\sigma_{V}$). The latter are
obtained as weighted means of the individual values deduced for each order in
V789~Mon and GZ~Leo spectra. We note that
 the uncertainties returned by {\sc fxcor} for SB2 binaries are overestimated;
 when fitting each component, the presence of the other component will increase the
 antisymmetric noise, thereby biasing the error.

\begin{figure}
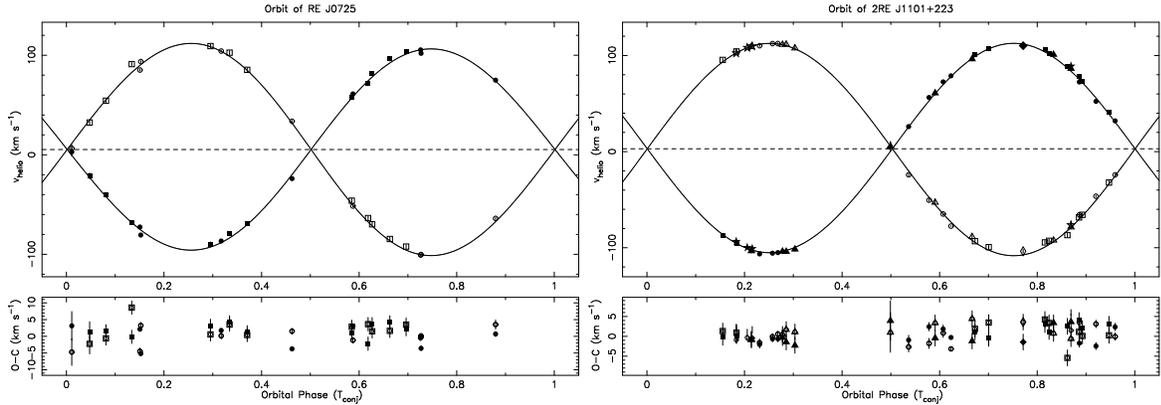

\begin{center}
\includegraphics[angle=270,width=7.6cm]{galvez_fig1.ps}
\includegraphics[angle=270,width=7.6cm]{galvez_fig2.ps}
\caption[ ]{Radial velocity data and fit vs. the orbital phase. 
Solid symbols represent the primary and open symbols represent the 
secondary. V789~Mon (2RE J0725-002) on the
left (squares data from Jeffries et al. (1995) and circles
data from our observations) and GZ~Leo (2RE J1101+223) on the rigth
 (squares data from Jeffries et al. 1995,
triangles from Fekel \& Henry 2000, rhombus from Strassmeier et
al. (2000), circles from our McDonald98 observing run and stars
from our FOCES02 observing run).
The solid curves represent a minimum $\chi^{2}$ fit orbit solution
as described in the text.
\label{fig:orb2n} }
\end{center}
\end{figure}

 To compute the orbital solution for these systems
 we combined our radial velocity data (for both components)
with the data of Jeffries et al. (1995),
Fekel $\&$ Henry (2000) and Strassmeier et al. (2000),
see Tables~\ref{tab:bvr} and~\ref{tab:vr}.
The radial velocity data are plotted in  Fig.~\ref{fig:orb2n}
 for V789~Mon and GZ~Leo on left and right side respectively. Solid
symbols represent the primary and open symbols represent the
secondary. The curves represent a minimum $\chi^{2}$ fit orbit
solution. The orbit fitting code uses the {\it Numerical Recipes}
 (Press et al. 1986)
implementation of the Levenberg-Marquardt method of fitting a
non-linear function to the data and weights each datum according
to its associated uncertainty (see e.g. G\'alvez et al. 2002, 
 2007 for further details).

We found a nearly circular ($e = 0.0129$) for V789~Mon system,
 with an orbital period of 1.4021 days which confirms it as a
synchronous system ($P_{\rm phot}$ $\approx$ 1.412 days, Cutispoto et al.
1999). The mass ratio ($q = 1.0542$) indicates that both components are
very similar. The obtained parameters are in agreement with
Jeffries et al. (1995). See Table~\ref{tab:orb2t} for details.

We arrived at a similar conclusions for GZ~Leo:
 a nearly circular orbit ($e = 0.0073$) with an orbital period of
 1.5260 days, which is very similar to 
 the photometric rotational period ($P_{\rm phot}$
$\approx$ 1.5264 days, Fekel \& Henry 2000), indicating a synchronized
rotation. The obtained minimum masses ($M\sin^{3}{i}$) and the mass
 ratio ($q = 1.0139$) are in agreement with Fekel \& Henry (2000).
 See Table~\ref{tab:orbt} for computed parameters.

\subsection{{\bf Physical Parameters}}
 Adopting spectral types derived in $\S$~3.1 for the primary components
 and adopting masses and temperatures from
Landolt-B\"{o}rnstein tables (Schmidt-Kaler 1982), we derived
 masses and spectral types for the secondaries.

We also estimated minimum radii ($R\sin{i}$), luminosities
 and minimum masses ($M\sin{i}$) of the components using photometric
 periods (from the literature, see $\S$~3.)
 and rotational velocities (from $\S$~ 3.2).
 The results, given in Table~\ref{tab:parb},
 confirm our previous classification.

 Furthermore, since $Hipparcos$ parallax of V789~Mon is not known, we
calculated a spectroscopic parallax.
Using bolometric magnitude of a K5V star (+6.7$^{m}$), the bolometric
 correction ($BC=-0.72^{m}$) corresponding to a K5V,
 and the luminosity ratio, we obtain the bolometric
and the visual total absolute magnitudes for the system.
 Comparing the total visual absolute magnitude (6.76$^{m}$),
 with the measured V magnitude (9.33$^{m}$), we derive a distance of
$32.72$~$\pm$0.13~pc ($\pi = 30.56$~$\pm$0.22~mas)
for V789~Mon. Due to possible
activity-induced variations in the V, we expect a 10\% additional
 error in this distance measure.
Since the system is relatively close, the interstellar reddening
is neglected.

Using the total luminosity based on the measured distance (54.26 pc)
for GZ~Leo, we obtained the total magnitude for the system $V=8.77$.
The difference between the observed $V$ magnitude (8.92) and
the calculated one can be explained by the presence of photospheric
cool spots over the stellar surfaces of both components.

\subsection{Kinematics}

We computed the galactic space-velocity components ($U$, $V$, $W$)
of these systems using as radial velocity the center of mass
 velocity ($\gamma$) and accurate proper motions and parallax taken
 from $Hipparcos$ (ESA 1997) and $Tycho-2$ (H$\o$g et al. 2000) catalogues
 (see Table~\ref{tab:par}), except for spectroscopic parallax of V789~Mon
 that has been calculated here.

The obtained values and associated errors are given in
 Table~\ref{tab:uvw}.
The velocity components lie in the  ($U$, $V$) diagram
 obviously inside the young disk population boundaries
(Eggen 1984, 1989; Montes et al. 2001a, 2001b) indicating
that both systems are young disk stars. In addition,
 the Eggen kinematic criteria predict that GZ~Leo
 may be a member of Hyades supercluster
 (see Montes et al. 2001a and  G\'alvez 2005).

\subsection{The Li~{\sc i} $\lambda$6707.8 line}

The resonance doublet of Li~{\sc i} at $\lambda$6707.8 \AA\ is a
spectroscopic feature very important in the diagnostic of
age in late-type stars, since it is destroyed easily by
 thermonuclear reactions in the stellar interior.

The spectral region of this line is included in all our spectra
 for both systems. 

\begin{figure}
\begin{center}
\includegraphics[width=7.0cm]{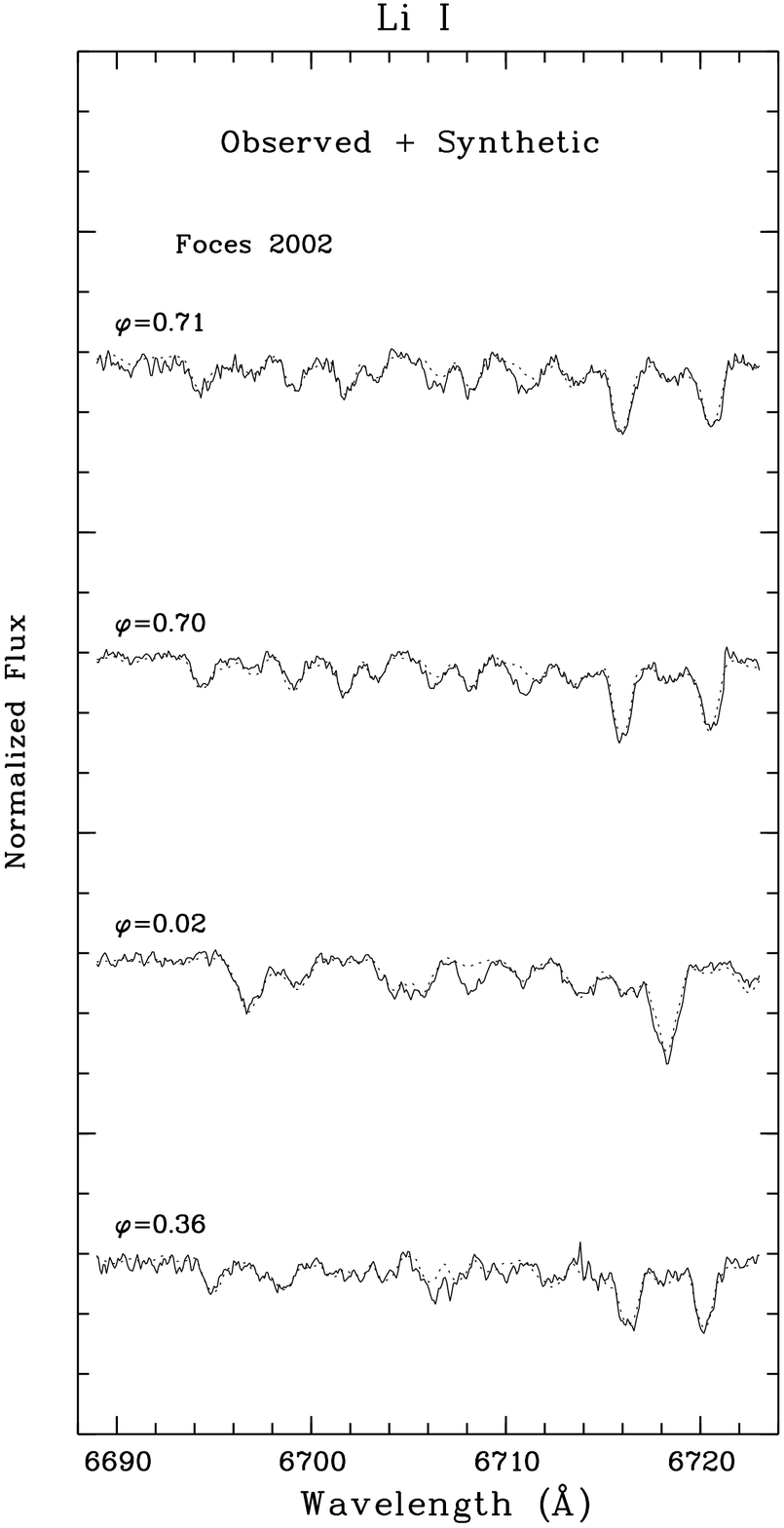}
\includegraphics[width=6.30cm]{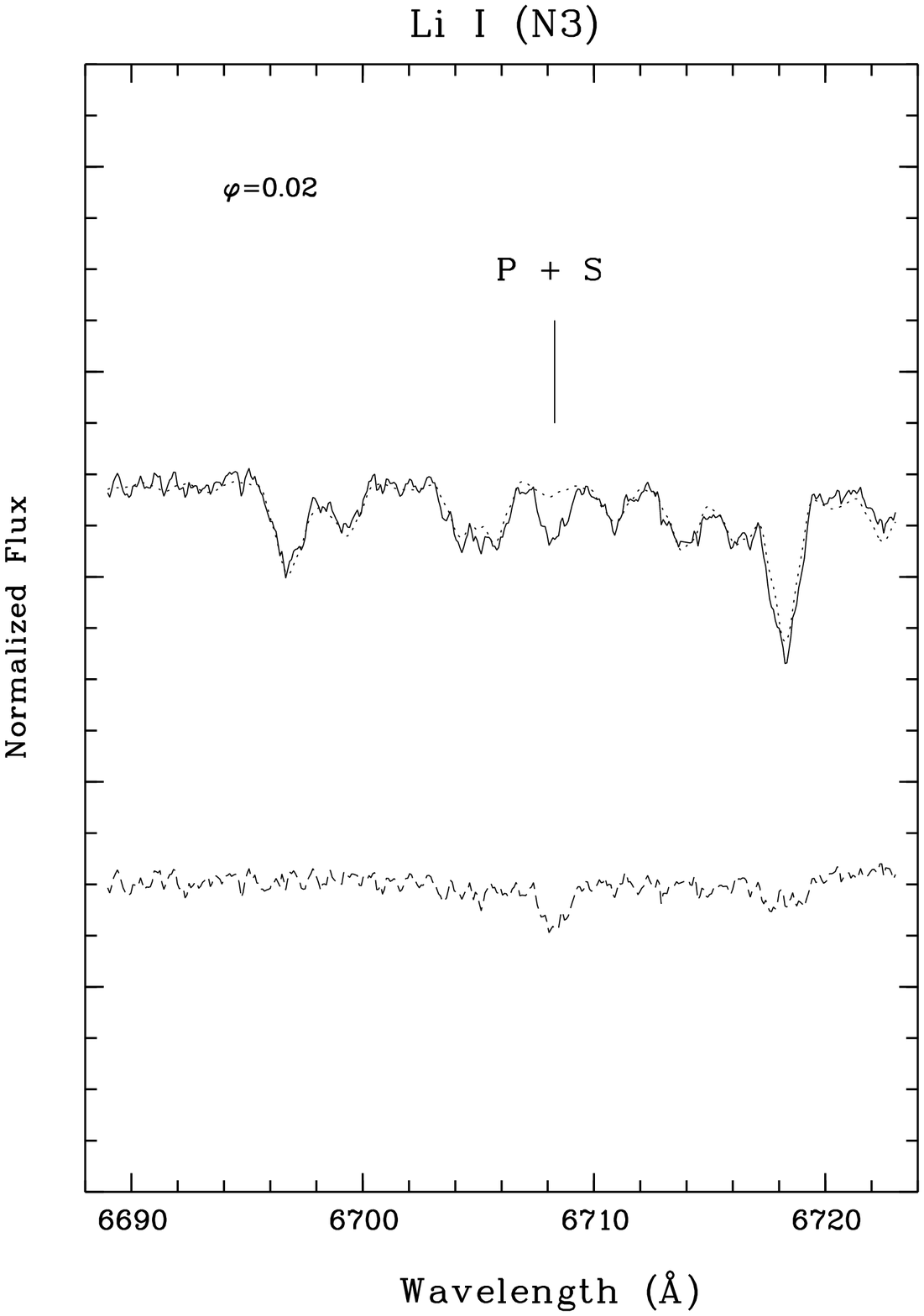}
\caption[ ]{On the left, we plot example spectra in different phases of 
 GZ~Leo in the Li~{\sc i} line region. 
 The observed (solid-line) and the
 synthesized spectrum (dashed-line) show the lithium
 absortion. On the right, we plot the conjuction phase 
 where we measured the $EW$(Li~{\sc i}).
The observed and the synthesized spectrum are plotted at the top and
 the subtracted spectrum (dotted line) at the bottom.
\label{fig:li}}
\end{center}
\end{figure}

 At the spectral resolution of our spectra and with the rotational velocity,
we have determined for the components of both binaries, the Li~{\sc i} line is blended with
the nearby Fe~{\sc i} $\lambda$6707.41~\AA\ line.
In the case of V789~Mon the Li~{\sc i} line is very weak 
and we could not measure the $EW$ with the
required precision. 
For GZ~Leo, we clearly see 
 the Li~{\sc i}+Fe~{\sc i}  absorption feature in all the spectra 
 (see Fig.~\ref{fig:li} left side), 
 but due to the blending with other photospheric
 lines of both components, we have only measured the $EW$ in the spectrum of 
 the third night of the FOCES02 observing run (see Fig.~\ref{fig:li} right) 
 which is very close to conjunction ($\varphi=0.02$).
 The obtained value $EW$(Li~{\sc i} + Fe~{\sc i}) = 60.7 m\AA\ contain the contribution from both components, 
 but taken into account that they have the same spectral type and contribution to the continuum 
 we can assume that this is also the $EW$(Li~{\sc i}+Fe~{\sc i}) of each component.
 We have corrected this total $EW$
by subtracting the $EW$(Fe~{\sc i}) = 19.8 m\AA\ calculated from the empirical
relationship with ($B$--$V$) given by Favata et al. (1993) resulting $EW$(Li~{\sc i}) = 40.9 m\AA . 
One additional estimation of the $EW$(Li~{\sc i}) in this spectrum can be obtained 
by using the spectral subtraction technique. 
In the subtracted spectrum the contribution from the photospheric lines is eliminated, therefore the 
$EW$ =  51.2 m\AA\ measured in this spectrum is the total $EW$(Li~{\sc i}) and, as both components 
are equal, also the $EW$(Li~{\sc i}) of each one.
 The difference in the $EW$ obtained from both methods is
 likely due to the influence of the stellar metallicity in the 
 calibration used in the first one to obtain the $EW$(Fe~{\sc i}) and in the standard star used 
 as reference in the subtraction technique that are not taken into
 account here.
 The $EW$(Li~{\sc i}) of each component we have determined (40.9 m\AA\ with the first method 
 and 51.2 m\AA\ with the second one)
  is smaller than the 63 m\AA\ estimated by Fekel \& Henry (2000)
 and the 60 m\AA\ given by Strassmeier et al. (2000).
 The small differences between these values can be attributed to the different method used by 
 these authors to estimate the Fe~{\sc i} contribution and the error in the $EW$ determination.

This small $EW$(Li~{\sc i})
 is in agreement with the values observed in Hyades cluster stars of this spectral type (see Montes et al. 2001b) 
 and with the membership to the Hyades supercluster (see $\S$~3.5). That situated the system age in
 about 600~Myr.

\section{Chromospheric activity indicators}
The echelle spectra analyzed in this work allowed us to
study the behavior of the different chromospheric  activity indicators, from
the Ca~{\sc ii} H \& K to the Ca~{\sc ii} IRT lines,
which are formed at different atmospheric heights.
The chromospheric contribution in these features were 
determined by using the spectral subtraction technique described in
detail by Montes et al. (1995, 1997, 1998, 2000) and
G\'alvez et al. (2002, 2007).

The excess emission in $EW$
is measured in the subtracted spectra,
after correcting for the relative contribution of
each component in the total continuum.
This contribution is determined by using the radii and
temperatures adopted in $\S$~3.
For instance, in the H$\alpha$ line region the relative contributions
were $S_{\rm P}=0.61$ for the primary component and
$S_{\rm S}=0.39$ for the secondary component in V789~Mon and for
GZ~Leo we had the 0.50 contributions for both components.

In Table~\ref{tab:ew} we give the corrected $EW$
 for the H$\alpha$ and Ca~{\sc ii} IRT
lines for V789~Mon. 
In Table~\ref{tab:ewf}
we list corrected $EW$ for the Ca~{\sc ii} H \& K, H$\epsilon$,
H$\delta$, H$\gamma$, H$\beta$, H$\alpha$, and  Ca~{\sc ii} IRT
($\lambda$8498, $\lambda$8542, $\lambda$8662 \AA) lines for GZ~Leo
in FOCES run and for
H$\alpha$ and Ca~{\sc ii} IRT in McDonald run.
 The corrected $EW$s is given separately for each component (P/S),
unless the lines of both components overlap, in which case
the combined $EW$ is given.
The final estimated errors in $EW$ are in the range 10-20\%
(see e.g. G\'alvez et al. 2007 for details).

Finally, corrected $EW$s were converted to
 the absolute surface fluxes by using the
empirical stellar flux scales calibrated by Hall (1996)
as a function of the star color index.
In our case, we used the $B-V$ index and the corresponding coefficients
for Ca~{\sc ii} H \& K, H$\alpha$ and Ca~{\sc ii} IRT.
 For H$\epsilon$ we used the same coefficients as for Ca~{\sc ii} H \& K,
while for H$\delta$, H$\gamma$ and H$\beta$ the coefficients
were obtained by interpolating 
between the values for Ca~{\sc ii} H \& K and H$\alpha$.
In Tables~\ref{tab:ew} and~\ref{tab:fluxf}, we give
 the logarithm of the calculated absolute flux at the stellar surface
(log$F$$_{\rm S}$) for different chromospheric activity indicators.

\begin{figure}
\begin{center}
\includegraphics[angle=270,width=8.3cm]{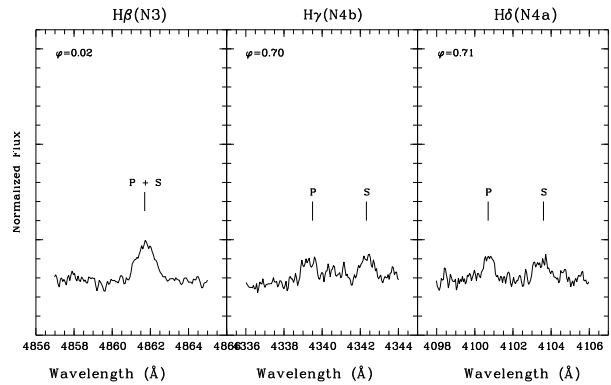}
\caption[ ]{Subtracted spectra in the region of the
H$\beta$, H$\gamma$, and H$\delta$ lines for GZ~Leo in FOCES observing run.
Clear excess emission from the primary (P) and secondary (S) components
is detected.
\label{fig:hbgd} }
\end{center}
\end{figure}

 Figs.~\ref{fig:ha3} - \ref{fig:irt}
 show a representative regions around the H$\alpha$, Ca~{\sc ii} H \& K  and
Ca~{\sc ii} IRT $\lambda$$8498$, $\lambda$$8542$ lines.
The observed (solid line)
and the synthesized spectra (dashed line) are shown in the left panel,
while subtracted spectrum (dotted line) in the right.
The observing run and the orbital phase ($\varphi$) of each spectrum
are also shown in these figures.
In Fig.~\ref{fig:hbgd}
we plot a representative subtracted spectra 
 in H$\beta$, H$\gamma$ and H$\delta$ region
for GZ~Leo.

\subsection{The H$\alpha$ line}

\begin{figure}
\begin{center}
\includegraphics[width=5.8cm]{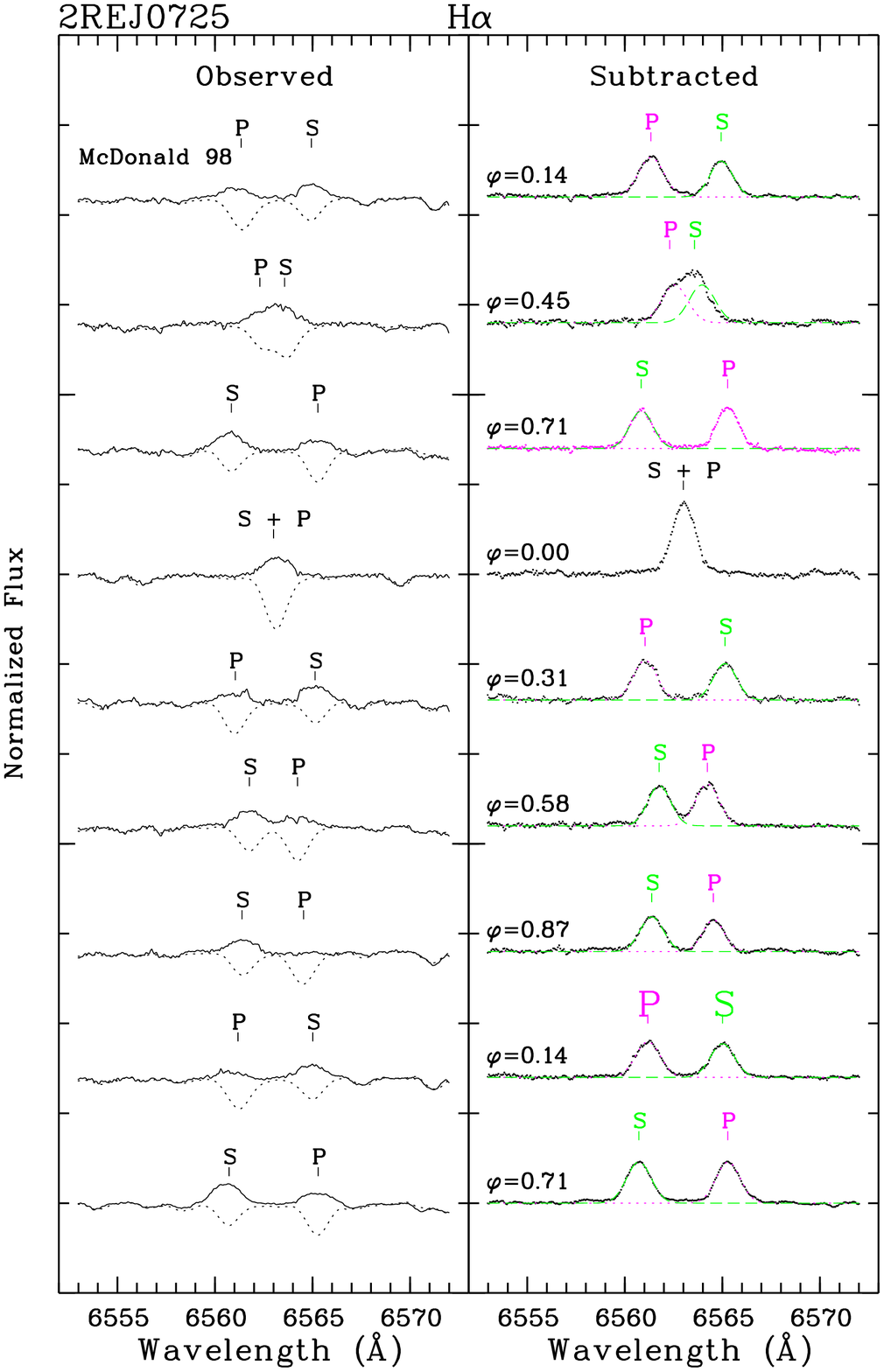}
\includegraphics[width=5.8cm]{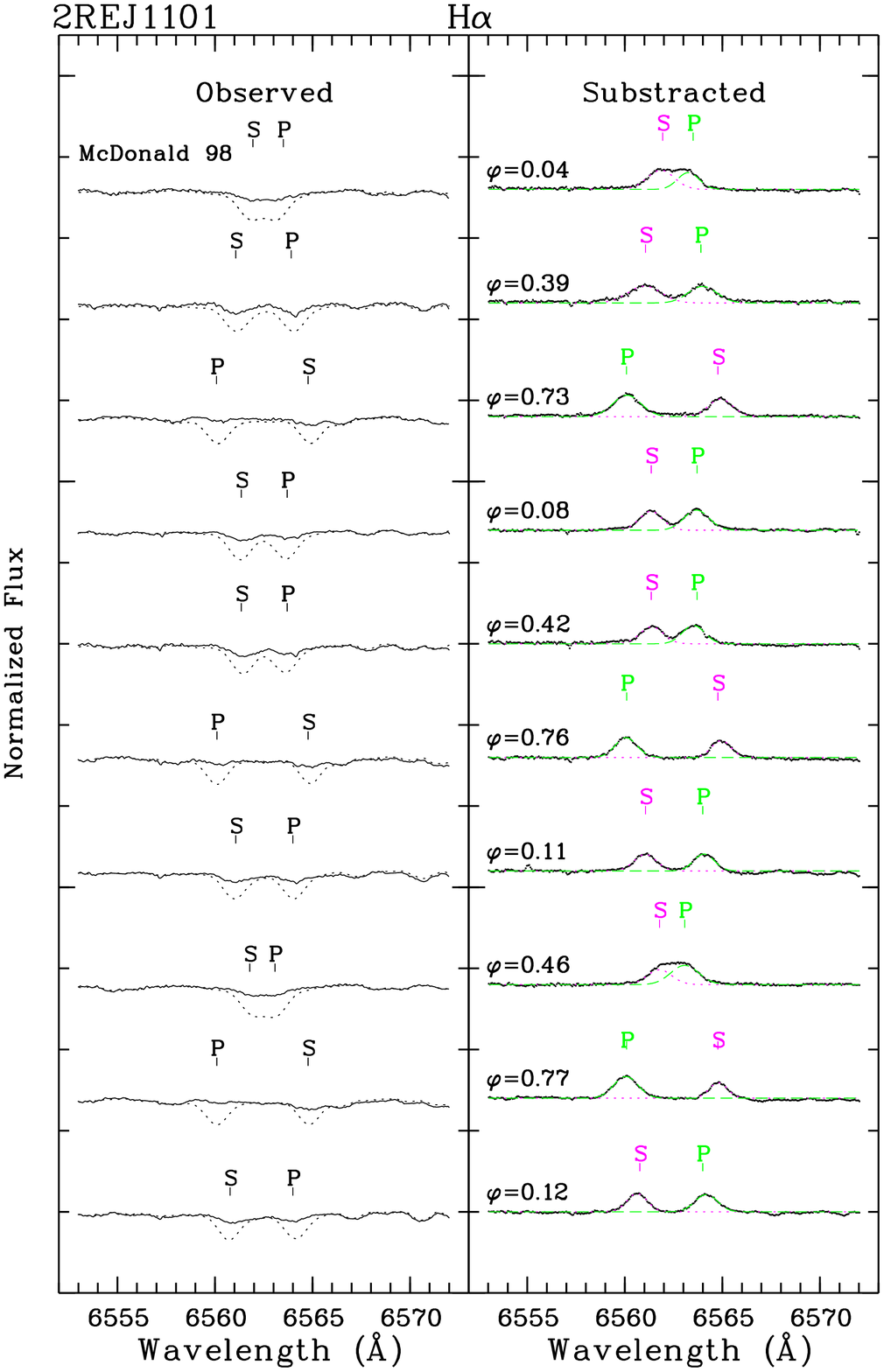}
\caption[ ]{Spectra of V789~Mon (2RE J0725) on the left and
 spectra of GZ~Leo (2RE J1101) on the right in the H$\alpha$ 
line region in McDonald run.
The observed spectrum (solid-line) and the
synthesized spectrum (dashed-line) are plotted in the left panel
and the subtracted spectrum (dotted line) in the right panel.
The position of the H$\alpha$ line for the primary (P) and secondary (S)
components are marked too and here we again have
the two-Gaussian fit superposed.
\label{fig:ha3}}
\end{center}
\end{figure}

In V789~Mon H$\alpha$ is
 observed in emission above the continuum
in all the observed spectra
(see Fig.~\ref{fig:ha3}, left panel of the left side).
After applying the spectral subtraction technique, 
one can see that H$\alpha$ emission
from the primary and secondary components looks very much alike.

In all the spectra, except one which is very close to conjunction,
we measured the emission coming from both components
by using a two-Gaussian fit to the subtracted spectra
(see Fig.~\ref{fig:ha3}, right panel of the left side).
The combined $EW$ is given
when it was not possible to deblend the two components.

The persistence of H$\alpha$ emission in V789~Mon spectra
indicates that it is a very active
 binary system similar to some RS CVn and BY Dra systems,
 which always show H$\alpha$ emission above the
continuum. The $EW$ measured in the subtracted spectra by Gaussian
fitting gives an average value of $EW$(H$\alpha$)~=~0.65/0.62 \AA\
  for primary and secondary
components (which is higher than reported by
 Jeffries et al. (1995), $EW$(H$\alpha$)~=~0.09/0.25 \AA).

In GZ~Leo, we analyzed 15 spectra including the H$\alpha$ line
 region. In the first run (McDonald98), the H$\alpha$ line appears
 completely filled by emission in both components
 (see Fig.~\ref{fig:ha3}, left panel of the right side).
As in the case of V789~Mon, the emission from both components
 is nearly identical and was measured by applying a
two-Gaussian fit to the subtracted spectra (see Fig.~\ref{fig:ha3}, right
 panel of the right side). 
 The spectrum at phase 0.46 is close to conjunction, so the
 measured $EW$ given in Table~\ref{tab:ewf} is the combined $EW$.
 The mean values of $EW$ are $EW_{P}$(H$\alpha$)~=~0.32 \AA\ and
$EW_{S}$(H$\alpha$)~=~0.36 \AA. 
 We note however that a variation with phase is
observed in both components, with a remarkable tendency
to behave in the opposite way for Ca~{\sc ii} IRT $EW$s;
unfortunately the phase coverage is not enough for
a reliable conclusion (Table~\ref{tab:ewf}).

When we analyze the FOCES02 run, we see again
that the H$\alpha$ line
is completely filled-in by emission for both components.
 After fitting the subtracted spectra we measured
 the excess emission $EW_{P}$(H$\alpha$)~=~0.46 \AA\ and
 $EW_{S}$(H$\alpha$)~=~0.47 \AA, i.e., a 35\% increase compared
 to McDonald98 run.

In GZ~Leo system the secondary star usually shows slightly larger
values of H$\alpha$ $EW$ and  no variation with phase was detected
 (see Table~\ref{tab:ewf}).

\subsection{The H$\beta$, H$\gamma$ and H$\delta$ lines}
The other three Balmer lines (H$\beta$, H$\gamma$ and H$\delta$)
 were included only in some of our spectra for GZ~Leo during the FOCES02.
 The absorption lines were clearly filled-in with emission in 
the observed spectra.
After applying the spectral subtraction, we could extract the excess
emission from both components (see representative spectra in
Fig.~\ref{fig:hbgd}). When the $S/N$ was high enough we
deblended the emission coming from both components by applying a
two-Gaussian fit to the subtracted spectra (see
Table~\ref{tab:ewf}).
These three lines show similar variations with orbital phase
 to the H$\alpha$ line
in both components, their mean values are
$EW$(H$\beta$)~=~0.13/0.13 \AA, $EW$(H$\gamma$)~=~0.09/0.09 \AA\
and $EW$(H$\delta$)~=~0.07/0.09 \AA.

We also measured the ratio of
excess emission $EW$ in the H$\alpha$ and H$\beta$ lines,
$EW({\rm H\alpha})/EW({\rm H\beta})$, and the ratio
of excess emission $E_{\rm H\alpha}/E_{\rm H\beta}$
with the correction:

\[ \frac{E_{\rm H\alpha}}{E_{\rm H\beta}} =
\frac{EW({\rm H\alpha})}{EW({\rm H\beta})}*0.2444*2.512^{(B-R)}\]
 from Hall \& Ramsey (1992). It takes into account
the absolute flux density in these lines and the color difference
in the components ($B-R$=1.45 for a K0V star).
We obtained a mean values of
$E_{\rm H\alpha}/E_{\rm H\beta}$ $\approx3.4$ for the primary and
 $\approx3.6$ for secondary.
These values indicate,
according to Buzasi (1989) and Hall \& Ramsey (1992)
the presence of prominence-like material at the stellar
surface in both components of the system.

\subsection{Ca~{\sc ii} H \& K and H$\epsilon$}

\begin{figure}
\begin{center}
\includegraphics[width=5.5cm]{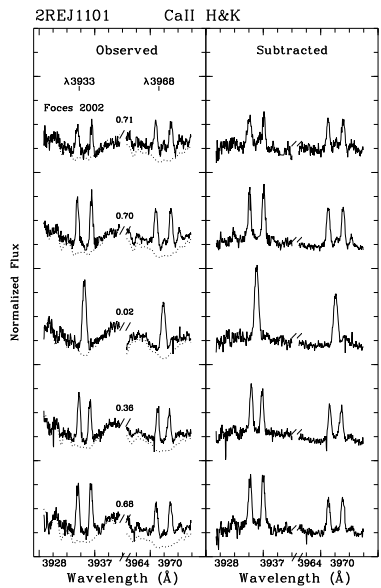}
\caption[ ]{As in Fig. 3 but in the region of the
Ca~{\sc ii} H \& K and H$\epsilon$ lines for GZ~Leo (2RE J1101) in FOCES run.
\label{fig:hyk} }
\end{center}
\end{figure}

The Ca~{\sc ii} H \& K line region was included only in the spectra
of the FOCES run for GZ~Leo.
In these spectra we observed strong emission in the Ca~{\sc ii} H \& K
lines and a clear emission in the H$\epsilon$ line
 arising from both components (see Fig.~\ref{fig:hyk}).

In our spectra, the Ca~{\sc ii} H \& K lines
 are located at the end of the echellogram, where the efficiency of
the spectrograph and the CCD decrease very rapidly, and therefore the
 $S/N$ ratio obtained is very low, and the normalization of the spectra
are very difficult. In spite of this we could apply the spectral
subtraction in this region, see
Fig.~\ref{fig:hyk}. As we can see in this figure, the
 H$\epsilon$ line arising from one of the component overlaps at
 some orbital phases with the Ca~{\sc ii} H line
 from the other component, so their $EW$ were measured with a
Gaussian fit when it was possible (see comments for each spectrum in
 the footnotes of Table~\ref{tab:ewf}).

The mean measured $EW$s in these spectra are
 $EW$=~1.13/0.84 \AA\ for each component in the Ca~{\sc ii}  K line
 and $EW$=~0.82/0.65 \AA\ for each component in the Ca~{\sc ii}  H line.
An increase of the Ca~{\sc ii} H \& K emissions is observed 
 with the orbital phase.

The presence of an excess emission from H$\epsilon$ line ($EW$=~0.30/0.43 \AA)
indicates that it is a very active system.

\subsection{Ca~{\sc ii} IRT lines
($\lambda$8498, $\lambda$8542, and $\lambda$8662)}
\begin{figure}
\begin{center}
\includegraphics[width=5.5cm]{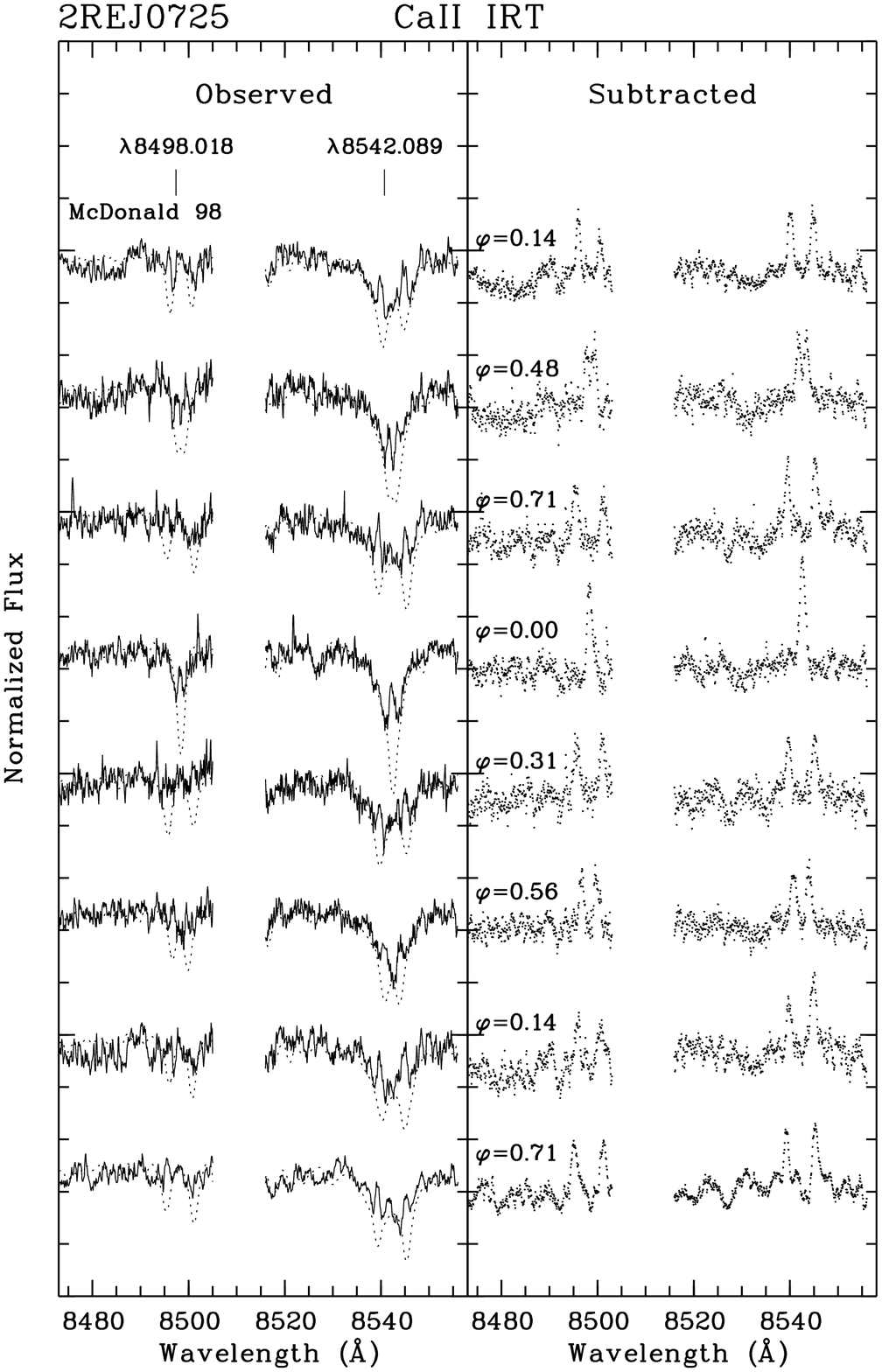}
\includegraphics[width=5.5cm]{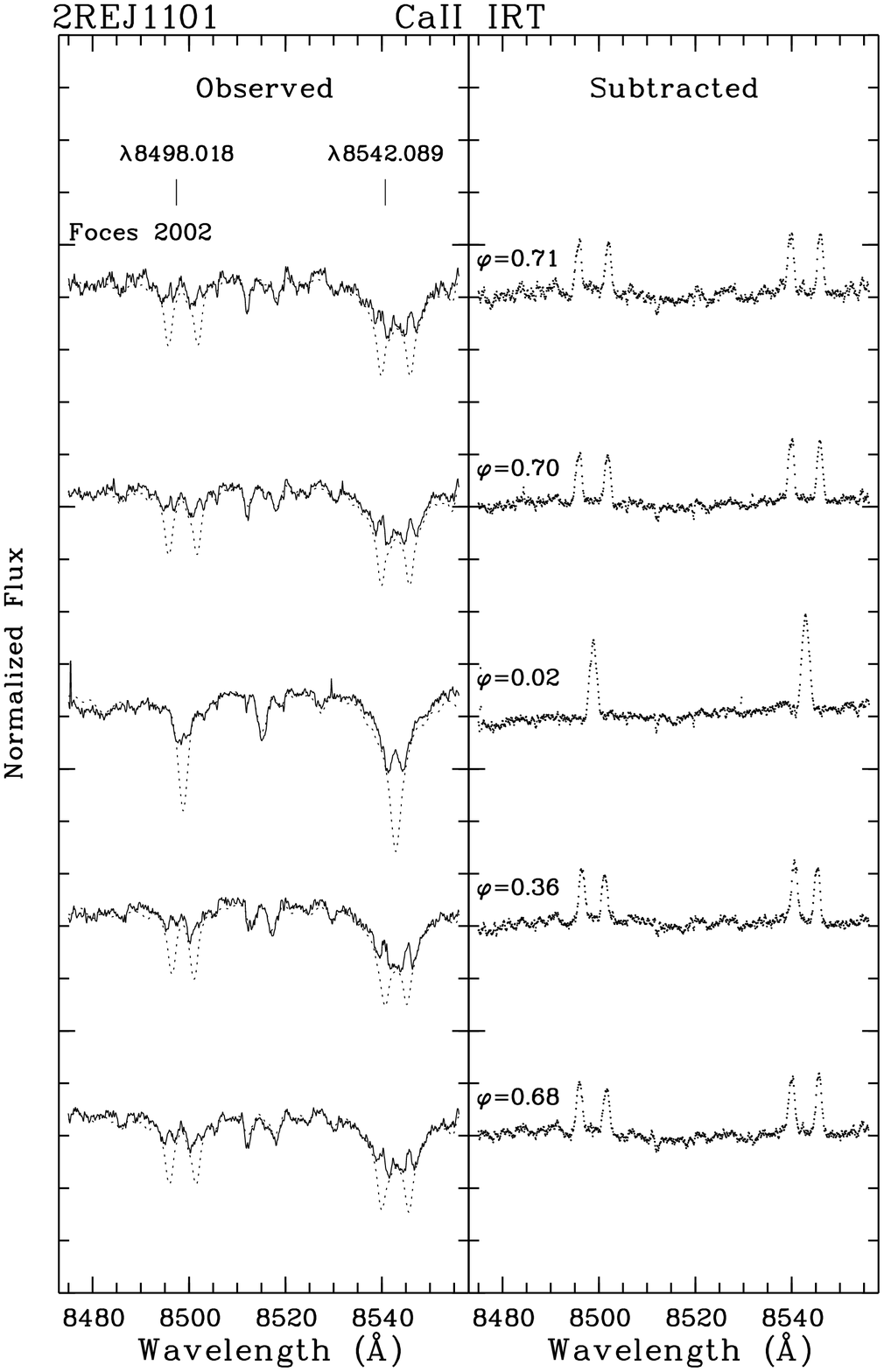}
\caption[ ]{The same as previous Figure but in the region of the
Ca~{\sc ii} IRT (8498, 8542~\AA) lines.
V789~Mon (2RE J0725) system on the left in McDonald run and
GZ~Leo (2RE J1101) on the right in FOCES run. 
 An excess emission from the primary (P) and secondary (S) components
is detected and marked.
\label{fig:irt} }
\end{center}
\end{figure}

The three lines of the Ca~{\sc ii} infrared triplet (IRT) were 
included in all our echelle spectra for both systems.
In all of them we could observe a clear emission above the continuum
 in the core of the Ca {\sc ii} IRT absorption lines
 (see Figs.~\ref{fig:irt}) from both components.

In V789~Mon, (Fig.~\ref{fig:irt} left side), we measured an average
$EW$ of $\approx0.4$ \AA\ for the three lines, and we found small
variations with the orbital phase that were anti-correlated with
 the Balmer lines $EW$ variations in both primary and secondary
components.

For GZ~Leo binary, (Fig.~\ref{fig:irt} right side),
we found that in FOCES run there were no significant variations
with the orbital phase, but in McDonald's one an evident orbital
variation were observed which shows an anti-correlation with
 the $EW$(H$\alpha$) variation.
 We  measured  mean
$EW$s of $\approx0.25/0.31$ \AA\ for each component in McDonald run
and $\approx0.36$ \AA\ for both components in FOCES run.

In addition, we calculated the ratio of excess emission $EW$,
$E_{8542}/E_{8498}$, which is also an indicator of the
type of chromospheric structure that produces the observed emission.
In solar plages, values of
$E_{8542}/E_{8498}$ $\approx$~1.5-3 are measured,
while in solar prominences the values are $\approx$~9,
the limit of an optically thin emitting plasma (Chester 1991).

In V789~Mon, small values of the $E_{8542}/E_{8498}$
ratio were found, $\approx$~1.5 for the primary component
and 1.4 for the secondary (see Table~~\ref{tab:ew}). This indicates 
that the Ca~{\sc ii} IRT emission of this star arises from
plage-like regions.

In GZ~Leo we found a $E_{8542}/E_{8498}$ value of
$\approx$~2.4 and $\approx$~1.4 for primary and secondary in McDonald run,
but $\approx$~0.9 for primary and secondary in
FOCES run.
These values indicate that Ca~{\sc ii} IRT emission come from plage-like
regions, in contrast to the Balmer lines which seem to come from prominences.

This markedly different behavior of the Ca~{\sc ii} IRT
 and the H$\alpha$ emission
has also been found in other chromospherically active binaries
 (see Ar\'evalo \& L\'{a}zaro 1999; Montes et al. 2000; G\'alvez 
 et al. 2003; G\'alvez 2005 and references therein). 
The different behavior from one epoch to another is consistent with
the different activity level measured in each observing run. 
 The lack of variation with orbital phase of the GZ~Leo emission lines
 in FOCES run could be due to that the hight level of activity is reflecting 
 a large number of active regions maybe distributed along
  all surface. Such, could produce an even spotted-plaged
 surface that would avoid to see the modulation clearly.

\section{Summary}
In this paper we present a detailed spectroscopic analysis of two 
X-ray/EUV selected chromospherically active binary systems
V789~Mon (2RE J0725-002) and GZ~Leo (2RE J1101+223).
We analyzed high resolution echelle spectra
that include the optical chromospheric activity indicators
from the Ca~{\sc ii} H \& K to Ca~{\sc ii} IRT lines, as well as the
 Li~{\sc i} $\lambda$6707.8 line and other photospheric lines of interest.

Using a large number of radial velocities we improved the orbital parameters of
these systems. We obtained that both systems have
 nearly circular orbits and have an orbital period very close to
 photometric period, indicating that they have a synchronous rotation.

The spectral classification derived using the comparison
with spectra of reference stars, two K5V components for V789~Mon
 and two K0V components for GZ~Leo is in good agreement with
 the classification obtained from physical parameters ($R\sin{i}$).
As both components are main sequence stars, we can classify these
chromospherically active binaries as BY$~$Dra systems
(Fekel et al. 1986).

By using the information provided by the width of the CCF we 
determined a projected rotational velocity, $v\sin{i}$, of 28.28 and 25.09
km\,s$^{-1}$ for the primary and secondary components of V789~Mon, and
26.23 and 26.98 km\,s$^{-1}$ for the components of GZ~Leo.

The presence of the Li~{\sc i} line in both systems is in agreement
with the kinematics results, i.e., they belong to the young disk population.
Also, with a $EW$(Li~{\sc i}) $\approx$ 40 m\AA\ for each
component, GZ~Leo could be a member of the Hyades supercluster.

We analyzed, using the spectral subtraction technique,
all the optical chromospheric activity indicators.
In V789~Mon, both components show
high levels of chromospheric activity. The emission in the H$\alpha$
line is above the continuum and it is very similar in the two components,
showing no variations with phase.
GZ~Leo shows a high level of chromospheric activity too.
After applying spectral subtraction the chromospheric excess emission
from both components are certainly detected in all the activity indicators
during the observing runs.
The Ca~{\sc ii} H \& K lines are in emission and H$\epsilon$ is also detected
in emission for both systems.

The level of chromospheric emission change, i.e., we have observed a 35\%
 increment in the $EW$ of H$\alpha$ emission line and a nearly 30\%
 increment in the $EW$ of Ca~{\sc ii} IRT emission lines from one
epoch to another.

Furthermore, the variation of H$\alpha$ and Ca~{\sc ii}~IRT
emission with phase is anti-correlated and the ratios $E_{\rm
H\alpha}/E_{\rm H\beta}$ and $E_{8542}/E_{8498}$ indicate
that the emission of the Balmer lines would arise from
prominence-like material, and that the emission of Ca~{\sc ii} IRT
lines arise from plage-like regions.


\acknowledgments
We would like to thank Dr. L.W. Ramsey (Pennsylvania State
University) for collaborating in the McDonald observing run (2.1~m
telescope, Texas, USA) and the staff of McDonald observatory for their
allocation of observing time and their assistance with our
observations. 
Thanks to Dr. Nadya Gorlova and
 Dr. Joern Rossa for their great help on the English redaction and
 corrections.
This work was supported the Spanish "Programa Nacional de
Astronom\'{\i}a y Astrof\'{\i}sica" under grants
AYA2005-02750 and AYA2008-00695, and the "Comunidad de Madrid"
under PRICIT project S-0505/ESP-0237 (ASTROCAM).



\clearpage

\onecolumn

\clearpage

\begin{table}
\caption[]{Observing runs
\label{tab:reduc1}}
\begin{flushleft}
\scriptsize
\begin{tabular}{llllcccc}
\noalign{\smallskip}
\hline \hline
\noalign{\smallskip}
 Date & Telescope & Instrument & CCD chip & Spect. range & Orders &
Dispersion & FWHM  \\
   &           &            &       & (\AA)   &          & (\AA/pixel) & (\AA) \\
\noalign{\smallskip}
\hline
\noalign{\smallskip}
12-21 01/1998  & 2.1m OS$^{\rm a}$ & Sandiford & 1200x400 CCD & 6400 - 8800 &32 & 0.06 - 0.08 & 0.13 - 0.20 \\
 22-25 04/2002    & 2.2m$^{\rm b}$  & FOCES       & 2048x2048 24$\mu$m Site\#1d
     & 3510 - 10700 & 112 & 0.04 - 0.13 & 0.10 - 0.40 \\
\noalign{\smallskip}
\hline
\noalign{\smallskip}
\end{tabular}

{
$^{\rm a}$ 2.1~m Otto Struve Telescope at McDonald Observatory (Texas, USA).\\
$^{\rm b}$ 2.2~m Telescope at the German Spanish Astronomical Observatory (CAHA) (Almer\'{\i}a, Spain).\\}

\end{flushleft}
\end{table}

\begin{table}
\caption[]{Stellar parameters
\label{tab:par}}
\begin{flushleft}
\scriptsize
\begin{tabular}{ll c c c  c l l l r r }
\noalign{\smallskip}
\hline \hline
\noalign{\smallskip}
{Name} &  {$T_{\rm sp}$} & {SB} &
 {$B-V$} & {$V$} & {$P_{\rm orb}^{*}$} & {$P_{\rm phot}$} &
{\it v}sin{\it i}$^{*}$ &
   $\pi$ & $\mu$$_{\alpha}$ cos $\delta$ & $    \mu$$_{\delta}$ \\
      &     &    &      &   &  (days) &  (days) &
(km\,s$^{-1}$) & (mas) & (mas/yr) & (mas/yr) \\
\noalign{\smallskip}
\hline
\noalign{\smallskip}
V789~Mon & K5V/K5V & 2 & 1.03  & 9.33 & 1.4021 & 1.412 & 28.28$\pm$1.59/25.09$\pm$2.06 & 30.56$\pm$0.22$^{*}$ & 73.20$\pm$1.00 & -27.10$\pm$1.00 \\
GZ~Leo & K1V/K1V & 2 & 0.87 & 8.92 & 1.5260 & 1.526 & 26.23$\pm$1.13/26.92$\pm$1.14 & 18.43$\pm$1.19 & -141.40$\pm$0.80 & 5.30$\pm$0.80 \\
\noalign{\smallskip}
\hline
\noalign{\smallskip}
\end{tabular}

{
 $^{*}$ values determined in this paper
\\
}
\end{flushleft}
\end{table}


\begin{table}
\caption[]{Radial velocities of V789~Mon
\label{tab:bvr}}
\begin{flushleft}
\scriptsize
\begin{tabular}{llrrrrrrrrrrrllrrrrrrrr}
\noalign{\smallskip}
\hline \hline
\noalign{\smallskip}
 Obs. & HJD & \tiny $S/N$ & \multicolumn{1}{c}{Primary} & \multicolumn{1}{c}{Secondary} \\
\noalign{\smallskip}
    & 2400000+& (H$\alpha$)& {\rm $V_{\rm hel}$} $\pm$ $\sigma_{\it V}$ &
            {\rm $V_{\rm hel}$} $\pm$ $\sigma_{\it V}$   \\
\noalign{\smallskip}
    &  & & \tiny (km\,s$^{-1}$) & \tiny (km\,s$^{-1}$)\\
\noalign{\smallskip}
\hline
\noalign{\smallskip}
 JEF(95)$^{*}$ & 49054.400 & - &  57.8 $\pm$ 2.0 & -46.1 $\pm$ 2.0 \\
 JEF(95)$^{*}$ & 49054.444 & - &  71.9 $\pm$ 2.0 & -63.6 $\pm$ 2.0 \\
 JEF(95)$^{*}$ & 49054.456 & - &  81.6 $\pm$ 2.0 & -69.8 $\pm$ 2.0 \\
 JEF(95)$^{*}$ & 49054.507 & - &  96.3 $\pm$ 2.0 & -84.6 $\pm$ 2.0 \\
 JEF(95)$^{*}$ & 49054.555 & - & 103.2 $\pm$ 2.0 & -92.2 $\pm$ 2.0 \\
 JEF(95)$^{*}$ & 49055.394 & - & -89.6 $\pm$ 2.0 & 109.2 $\pm$ 2.0 \\
 JEF(95)$^{*}$ & 49055.451 & - & -79.5 $\pm$ 2.0 & 102.6 $\pm$ 2.0 \\
 JEF(95)$^{*}$ & 49055.500 & - & -69.1 $\pm$ 2.0 &  85.4 $\pm$ 2.0 \\
 JEF(95)$^{*}$ & 49056.449 & - & -21.2 $\pm$ 2.0 &  32.4 $\pm$ 2.0 \\
 JEF(95)$^{*}$ & 49056.496 & - & -40.3 $\pm$ 2.0 &  54.3 $\pm$ 2.0 \\
 JEF(95)$^{*}$ & 49056.572 & - & -68.1 $\pm$ 2.0 &  91.1 $\pm$ 2.0 \\
 MCD98 & 50826.835  & 152 & 105.40 $\pm$ 0.66 & -100.66 $\pm$0.88  \\
 MCD98 & 50828.837  & 76 & -80.61 $\pm$ 0.66 & 93.65 $\pm$0.93
 \\
 MCD98 & 50829.854  & 70 & 74.97 $\pm$ 0.76 & -63.90 $\pm$1.38
 \\
 MCD98 & 50830.848  & 59 & 61.17 $\pm$ 0.67 & -51.52 $\pm$0.86 \\
 MCD98 & 50831.870  & 46 & -86.62 $\pm$ 0.62 & 104.17 $\pm$0.95 \\
 MCD98 & 50833.846  & 58 & 101.92 $\pm$ 0.67 & -100.26 $\pm$0.83 \\
 MCD98 & 50834.879  & 56 & -23.90 $\pm$ 0.62 & 33.66 $\pm$0.89
 \\
 MCD98 & 50835.844  & 47 & -72.60 $\pm$ 0.66 & 85.18 $\pm$1.00 \\
 MCD98 & 50832.843  & 99 & 2.87 $\pm$ 4.28 &  6.44 $\pm$4.00
\\
\noalign{\smallskip}
\hline
\noalign{\smallskip}
\end{tabular}

{
 $^{*}$ JEF(95): Jeffries et al. (1995)}

\end{flushleft}
\end{table}

\begin{table}
\caption[]{Radial velocities of GZ~Leo
\label{tab:vr}}
\begin{flushleft}
\scriptsize
\begin{tabular}{llrrrrrrrrrrrllrrrrrrrr}
\noalign{\smallskip}
\hline \hline
\noalign{\smallskip}
 Obs. & HJD & \tiny $S/N$ & \multicolumn{1}{c}{Primary} & \multicolumn{1}{c}{Secondary} \\
\noalign{\smallskip}
    & 2400000+& (H$\alpha$)& {\rm $V_{\rm hel}$} $\pm$ $\sigma_{\it V}$ &
            {\rm $V_{\rm hel}$} $\pm$ $\sigma_{\it V}$   \\
\noalign{\smallskip}
    &  & & (km\,s$^{-1}$) & (km\,s$^{-1}$)\\
\noalign{\smallskip}
\hline
\noalign{\smallskip}
 JEF(95)$^{a}$ & 49052.571 & - & 101.9 $\pm$ 2.0 & -92.8 $\pm$ 2.0\\
 JEF(95)$^{a}$ & 49052.664 & - &  78.5 $\pm$ 2.0 & -68.5 $\pm$ 2.0\\
 JEF(95)$^{a}$ & 49052.757 & - &  40.9 $\pm$ 2.0 & -32.2 $\pm$ 2.0\\
 JEF(95)$^{a}$ & 49054.600 & - & -87.2 $\pm$ 2.0 &  95.4 $\pm$ 2.0\\
 JEF(95)$^{a}$ & 49054.644 & - & -95.9 $\pm$ 2.0 & 104.2 $\pm$ 2.0\\
 JEF(95)$^{a}$ & 49055.389 & - & 100.7 $\pm$ 2.0 & -93.2 $\pm$ 2.0\\
 JEF(95)$^{a}$ & 49055.435 & - & 106.8 $\pm$ 2.0 & -99.4 $\pm$ 2.0\\
 JEF(95)$^{a}$ & 49055.611 & - & 106.2 $\pm$ 2.0 & -94.5 $\pm$ 2.0\\
 JEF(95)$^{a}$ & 49055.679 & - &  88.7 $\pm$ 2.0 & -86.9 $\pm$ 2.0\\
 JEF(95)$^{a}$ & 49055.727 & - &  72.8 $\pm$ 2.0 & -65.8 $\pm$ 2.0\\
 FEKEL(00)$^{b}$ & 50200.709 & - & -103.4 $\pm$ 2.0 &  109.1 $\pm$ 2.0\\
 FEKEL(00)$^{b}$ & 50400.035 & - &  101.0 $\pm$ 2.0 &  -92.3 $\pm$ 2.0\\
 FEKEL(00)$^{b}$ & 50401.049 & - &    5.4 $\pm$ 2.0 &    5.4 $\pm$ 2.0\\
 FEKEL(00)$^{b}$ & 50576.681 & - &   60.9 $\pm$ 2.0 &  -52.9 $\pm$ 2.0\\
 FEKEL(00)$^{b}$ & 50632.664 & - & -103.9 $\pm$ 2.0 &  111.4 $\pm$ 2.0\\
 FEKEL(00)$^{b}$ & 50831.054 & - & -104.0 $\pm$ 2.0 &  111.6 $\pm$ 2.0\\
 FEKEL(00)$^{b}$ & 50831.947 & - &   86.4 $\pm$ 2.0 &  -78.6 $\pm$ 2.0\\
 FEKEL(00)$^{b}$ & 50927.773 & - &   96.4 $\pm$ 2.0 &  -88.6 $\pm$ 2.0\\
 MCD98 & 50826.992 & 123 & -77.48 $\pm$ 0.44 & 78.84 $\pm$ 0.16 \\
 MCD98 & 50827.975 & 133 & 112.42 $\pm$ 0.75 & -105.11 $\pm$ 0.78 \\
 MCD98 & 50829.032 & 123 & -24.19 $\pm$ 0.93 & 31.98 $\pm$ 0.96 \\
 MCD98 & 50830.019 & 116 & -64.94 $\pm$ 1.10 & 72.53 $\pm$ 1.02 \\
 MCD98 & 50831.010 & 86 &  112.36 $\pm$ 0.77 & -105.72 $\pm$ 0.76 \\
 MCD98 & 50832.024 & 92 & -46.47 $\pm$ 0.96 & 52.29 $\pm$ 0.90 \\
 MCD98 & 50833.027 & 132 & -50.52 $\pm$ 1.28 & 56.33 $\pm$ 1.05 \\
 MCD98 & 50834.024 & 100 & 110.18 $\pm$ 0.92 & -106.47 $\pm$ 0.92 \\
 MCD98 & 50835.025 & 80 & -65.73 $\pm$ 1.12 & 72.44 $\pm$ 1.01 \\
 MCD98 & 50836.017 & 97 & -24.11 $\pm$ 1.26 & 26.04 $\pm$ 1.20 \\
 FEKEL(00)$^{b}$ & 51240.763 & - &  110.2 $\pm$ 2.0 & -103.7 $\pm$ 2.0\\
 STRAS(00)$^{c}$ & 51659.696 & - & -101.6 $\pm$ 2.0 & 107.7 $\pm$ 2.0\\
 FOCES02 & 52387.409 & 77 & -93.69 $\pm$ 0.75 & 102.35 $\pm$ 0.78 \\
 FOCES02 & 52388.458 & 82 & 88.72 $\pm$ 0.76 & -76.47 $\pm$ 0.96 \\
 FOCES02 & 52389.460 & 97 & 4.76 $\pm$ 0.24 & 4.76 $\pm$ 0.24 \\
 FOCES02 & 52390.496 & 96 & -99.96 $\pm$ 0.69 & 108.04 $\pm$ 0.79 \\
 FOCES02 & 52390.512 & 68 & -101.09 $\pm$ 082 & 109.61 $\pm$ 0.68 \\
\\
\noalign{\smallskip}
\hline
\noalign{\smallskip}
\end{tabular}

{
$^{a}$ JEF(95): Jeffries et al. (1995)
\\
$^{b}$ FEKEL(00): Fekel  $\&$ Henry (2000)
\\
$^{c}$ STRAS(00): Strassmeier et al. (2000)}

\end{flushleft}
\end{table}

\begin{table}
\caption[]{Orbital solution of V789~Mon
\label{tab:orb2t}}
\scriptsize      
\begin{flushleft}
\begin{tabular}{lccc}
\noalign{\smallskip}
\hline \hline
\noalign{\smallskip}
Element & Value & Uncertainty & Units \\
\noalign{\smallskip}
\hline
\noalign{\smallskip}
 $P_{\rm orb}$ & 1.4021     & 0.0000  &  days  \\
 $T_{\rm conj}$ & 49054.980   & 0.048  & HJD  \\
 $\omega$ & 269.44   & 0.23  & degrees  \\
 $e$        &     0.0129 & 0.0024  &  \\
 $K_{\rm P}$  &   101.09 & 0.47  & km\,s$^{-1}$\\
 $K_{\rm S}$  &   106.57 & 0.69  &  km\,s$^{-1}$  \\
 $\gamma$ &     5.29 & 0.17  & km\,s$^{-1}$  \\
 $q=M_{\rm P}/M_{\rm S}$    &    1.0542  & 0.0047  &   \\
\\
 $a_{\rm P}$~sin$i$  &     1.9490 & 0.0092  & 10$^{6}$~km \\
 $a_{\rm S}~$sin$i$  &    2.054  & 0.013  &  10$^{6}$~km \\
 $a$~sen$i$       &     4.003 & 0.016  &  10$^{6}$~km \\
 "            &     0.027&         &  AU  \\
 "            &     5.752 &         &  R$_{\odot}$ \\
\\
 $M_{\rm P}$~sin$^{3}i$  &    0.6675  & 0.0085  & \small $M_{\odot}$\\
 $M_{\rm S}$~sin$^{3}i$  &    0.6332  & 0.0081  & \small $M_{\odot}$ \\
 f($M$)     &   0.1500& 0.0021 & \small $M_{\odot}$ \\
\noalign{\smallskip}
\hline
\noalign{\smallskip}
\end{tabular}
\end{flushleft}
\end{table}

\begin{table}
\caption[]{Orbital solution of GZ~Leo
\label{tab:orbt}}
\scriptsize
\begin{flushleft}
\begin{tabular}{lccc}
\noalign{\smallskip}
\hline \hline
\noalign{\smallskip}
Element & Value & Uncertainty & Units \\
\noalign{\smallskip}
\hline
\noalign{\smallskip}
 $P_{\rm orb}$ & 1.5260     & 0.0000  &  days  \\
 $T_{\rm conj}$ & 49051.312   & 0.080  & HJD  \\
 $\omega$ &  359.33  & 18.98 & degrees  \\
 $e$        &     0.0073 & 0.0025  &  \\
 $K_{\rm P}$  &   108.92 & 0.48  & km\,s$^{-1}$\\
 $K_{\rm S}$  &   110.43 & 0.58  &  km\,s$^{-1}$  \\
 $\gamma$ &     2.93 & 0.12  &  km\,s$^{-1}$  \\
 $q=M_{\rm P}/M_{\rm S}$    &    1.0139  & 0.0029  &   \\
\\
 $a_{\rm P}$~sin$i$  &     2.285 & 0.010  & 10$^{6}$~km \\
 $a_{\rm S}~$sin$i$  &    2.317  & 0.012  &  10$^{6}$~km \\
 $a$~sen$i$       &     4.603 & 0.016  & 10$^{6}$~km \\
 "            &     0.031 &         & AU  \\
 "            &     6.613 &         & R$_{\odot}$ \\
\\
 $M_{\rm P}$~sin$^{3}i$  &    0.8400  & 0.0091  & $M_{\odot}$\\
 $M_{\rm S}$~sin$^{3}i$  &    0.8285  & 0.0090  &  $M_{\odot}$ \\
 f($M$)     &   0.2043 & 0.0027 & $M_{\odot}$ \\
\noalign{\smallskip}
\hline
\noalign{\smallskip}
\end{tabular}
\end{flushleft}
\end{table}

\begin{table}
\caption[]{Derived stellar parameters
\label{tab:parb}}
\begin{flushleft}
\scriptsize
\begin{tabular}{llcclccccccc}
\noalign{\smallskip}
\hline \hline
\noalign{\smallskip}
Name & ${SpT_{\rm P}}^{a}$ &  ${M_{\rm P}}^{a}$ & $M_{\rm S}$ &
 ${SpT_{\rm S}}^{b}$ &
  $R\sin{i}_{\rm P}$ & $R\sin{i}_{\rm S}$ & $T_{\rm eff}^{a}$ &
   $L_{\rm P}$ & $L_{\rm S}$ & $M_{\rm Pmin}$ & $M_{\rm Smin}$ \\
 &     &  ($M_{\odot}$)  & ($M_{\odot}$)  &      & ($R_{\odot}$)  & ($R_{\odot}$) &
 (K) & ($L_{\odot}$) & ($L_{\odot}$) & ($M_{\odot}$) & ($M_{\odot}$) \\
\noalign{\smallskip}
\hline
\noalign{\smallskip}
V789~Mon & K5V & 0.67 & 0.63 & K6-K7V & 0.79$\pm$0.03 & 0.70$\pm$0.06 & 4350 & 0.20 & 0.16 & 0.56 & 0.52 \\
GZ~Leo & K0V & 0.79 & 0.78 & K0V & 0.79$\pm$0.03 & 0.82$\pm$0.03 & 5250 & 0.42 & 0.54 & 0.73 & 0.75 \\
\noalign{\smallskip}
\hline
\noalign{\smallskip}
\end{tabular}
{\small
\\
 $^{a}$ From Landolt-B\"{o}rnstein tables (Schmidt-Kaler 1982)
\\
 $^{b}$ Derived
}
\end{flushleft}
\end{table}

\begin{table}
\caption[]{Galactic space-velocity components
\label{tab:uvw}}
\begin{flushleft}
\scriptsize
\begin{tabular}{lllll}
\noalign{\smallskip}
\hline \hline
\noalign{\smallskip}
Name & $U\pm \sigma_{\it U}$ & $V \pm \sigma_{\it V}$ & $W \pm \sigma_{\it W}$ & $V_{\rm Total}$ \\
           &(km\,s$^{-1}$) & (km\,s$^{-1}$) & (km\,s$^{-1}$)  & (km\,s$^{-1}$) \\
\noalign{\smallskip}
\hline
\noalign{\smallskip}
V789~Mon & -32.78$\pm$2.07 & -10.67$\pm$0.70 & -12.10$\pm$0.96 & 36.53 \\
GZ~Leo & 0.33$\pm$0.46 & -7.92$\pm$0.49 & 6.65$\pm$0.60 & 10.37 \\
\noalign{\smallskip}
\hline
\noalign{\smallskip}
\end{tabular}
\end{flushleft}
\end{table}

\begin{table}
\caption[]{$EW$ and absolute surface flux of the different chromospheric activity indicators of V789~Mon
\label{tab:ew}}
\begin{flushleft}
\scriptsize
\begin{tabular}{ccccccccccccccccccc}
\noalign{\smallskip}
\hline \hline
\noalign{\smallskip}
$\varphi$ &
\multicolumn{4}{c}{$EW$(\AA) (Primary/Secondary)} &\multicolumn{4}{c}{log$F$$_{\rm S}$ (Primary/Secondary)}\\
\cline{2-5}\cline{6-9}
\noalign{\smallskip}
     & &
      \multicolumn{3}{c}{Ca~{\sc ii} IRT} & & \multicolumn{3}{c}{Ca~{\sc ii} IRT} \\
\cline{3-5}\cline{7-9}
\noalign{\smallskip}
   & H$\alpha$ &
$\lambda$8498 & $\lambda$8542 & $\lambda$8662& H$\alpha$ &
$\lambda$8498 & $\lambda$8542 & $\lambda$8662
\scriptsize
\\
\noalign{\smallskip}
\hline
\noalign{\smallskip}
0.71 & 0.71/0.72 & 0.27/0.31 & 0.42/0.29 & 0.33/0.29 & 6.83/7.02 & 6.27/6.45 & 6.46/6.42 & 6.35/6.42 \\
0.14 & 0.62/0.58 & 0.19/0.22 & 0.33/0.53 & 0.13/0.36 & 6.77/6.93 & 6.11/6.23 & 6.35/6.68 & 5.95/6.51 \\
0.87 & 0.48/0.58 & - & - & - & 6.66/6.93 & - & - & - \\
0.58 & 0.70/0.65 & 0.30/0.25 & 0.34/0.39 & 0.34/0.28 & 6.82/6.98 & 6.31/6.35 & 6.36/6.55 & 6.36/6.40 \\
0.31 & 0.66/0.60 & 0.35/0.34 & 0.32/0.33 & 0.38/0.37 & 6.80/6.94 & 6.38/6.49 & 6.34/6.47 & 6.41/6.52 \\
0.00 & 1.24$^{a}$ &
0.39$^{a}$ & 0.63$^{a}$ & 0.52$^{a}$ & 7.07$^{a}$ & 6.42$^{a}$ & 6.63$^{a}$ & 6.55$^{a}$ \\
0.71 & 0.64/0.62 & 0.19/0.28 & 0.43/0.36 & 0.28/0.28 & 6.78/6.96 & 6.11/6.40 & 6.47/6.51 & 6.28/6.40 \\
0.45 &  1.39$^{a}$ & 0.23$^{b}$/0.41$^{b}$ & 0.28$^{b}$/0.30$^{b}$ & 0.28$^{b}$/0.40$^{b}$ & 7.12$^{a}$& 6.20$^{b}$/6.60$^{b}$ & 6.28$^{b}$/6.43$^{b}$ & 6.28$^{b}$/6.56$^{b}$ \\
0.14 & 0.72/0.58 & 0.25/0.18 & 0.36/0.34 & 0.24/0.30 & 6.83/6.93 & 6.23/6.21 & 6.39/6.49 & 6.21/6.43 \\
\noalign{\smallskip}
\hline
\noalign{\smallskip}
\end{tabular}

{
$\varphi$ Orbital phase (see Table~\ref{tab:reduc1})\\
 $^{a}$ Data for the primary and secondary components not
deblended.
\\
$^{b}$ Data measured with low $S/N$.}
\end{flushleft}
\end{table}

\begin{table}
\caption[]{$EW$ of the different chromospheric activity indicators of GZ~Leo 
\label{tab:ewf}}
\begin{flushleft}
\scriptsize
\begin{tabular}{lccccccccccc}
\noalign{\smallskip}
\hline \hline
\noalign{\smallskip}
 &          & \multicolumn{10}{c}{$EW$(\AA) in the subtracted spectrum (P/S)} \\
\cline{3-12}
\noalign{\smallskip}
O &  $\varphi$ & \multicolumn{2}{c}{Ca~{\sc ii}} & & & & & &
\multicolumn{3}{c}{Ca~{\sc ii} IRT} \\
\cline{3-4}\cline{10-12}
\noalign{\smallskip}
  &   &
 K   & H  & H$\epsilon$ & H$\delta$ & H$\gamma$ & H$\beta$ & H$\alpha$ &
\tiny       $\lambda$8498 & \tiny       $\lambda$8542 & \tiny       $\lambda$8662
\tiny      
\\
\noalign{\smallskip}
\hline
\noalign{\smallskip}
M & 0.12 & - & - & - & - & - & - & 0.29/0.30 & 0.30/0.20 & 0.22/0.21 & 0.26/0.26 \\
M & 0.77 & - & - & - & - & - & - & 0.22/0.41 & 0.23/0.33 & 0.41/0.52 & 0.37/0.40 \\
M & 0.46 & - & - & - & - & - & - & 0.29/0.42 & $^{c}$/0.49 & 0.90$^{a}$/$^{c}$ & 0.54$^{a}$/$^{c}$ \\
M & 0.11 & - & - & - & - & - & - & 0.29/0.28 & - & - & - \\
M & 0.76 & - & - & - & - & - & - & 0.26/0.34 & 0.13/0.39 & 0.33/0.29 \\
M & 0.42 & - & - & - & - & - & - & 0.31/0.32 &0.21/0.25 & 0.29/0.25 & 0.27/0.34 \\  
M & 0.08 & - & - & - & - & - & - & 0.34/0.40 & 0.14/0.18 & 0.35/0.30 & $^{c}$/$^{c}$ \\
M & 0.73 & - & - & - & - & - & - & 0.33/0.47 & 0.13/0.28 & 0.32/0.37 & 0.24/0.23 \\
M & 0.39 & - & - & - & - & - & - & 0.44/0.38 & 0.23/0.20 & 0.37/0.38 & 0.16/0.24 \\
M & 0.04 & - & - & - & - & - & - & 0.41/0.28 & 0.18/0.32 & 0.27/0.38 & 0.22/0.41 \\
F & 0.68 & 1.06/1.02 & 0.85/0.92$^{e}$ & $^{d}$/0.50 & 0.09/0.13$^{b}$ & $^{c}$/0.08$^{b}$ & 0.12/0.08$^{b}$ & 0.54/0.48 & 0.33/0.35 & 0.33/0.30 & 0.37/0.35 \\
F & 0.36 & 0.77/0.84 & 0.80$^{e}$/0.68 & 0.17/$^{d}$ & 0.05/0.09$^{b}$ & 0.08/0.10$^{b}$ & 0.09/0.15$^{b}$ & 0.32/0.52 & 0.39/0.40 & 0.33/0.37 & 0.34/0.42 \\
F & 0.02 & 1.52$^{a}$ & 1.27$^{a}$ & 0.68$^{a}$ & 0.10$^{a,}$$^{b}$ & 0.10$^{a,}$$^{b}$ & 0.21$^{a,}$$^{b}$ & 0.74$^{a}$ & 0.60$^{a}$ & 0.50$^{a}$ & 0.78$^{a}$ \\
F & 0.70 & 0.92/0.87 & 0.92/0.59 & 0.20/0.35 & 0.07/0.08$^{b}$ & 0.12/0.09$^{b}$ & 0.18/0.16$^{b}$ & 0.49/0.43 & 0.36/0.37 & 0.33/0.32 & 0.41/0.38 \\
F & 0.71 & 0.76/0.63 & 0.70/0.68 & 0.40/0.44 & 0.08/0.08$^{b}$ & 0.08/0.09$^{b}$ & 0.13/0.13$^{b}$ & 0.50/0.46 & 0.36/0.37 & 0.35/0.32 & 0.40/0.37 \\
\noalign{\smallskip}
\hline
\noalign{\smallskip}
\end{tabular}

{
%
O Observing run: M: McDonald98, F: FOCES02.
\\
$\varphi$ Orbital phase (see Table~\ref{tab:reduc1})\\
$^{a}$ Data for the primary and secondary components not deblended.
\\
$^{b}$ Mean value of measured (from two apertures in our spectra)
 or value from higher $S/N$ aperture \\
$^{c}$ Data not measured due to very low $S/N$.
\\
$^{d}$ Data not measured due to blending.
\\
$^{e}$ Ca~{\sc ii} H line of one component blended with H$\epsilon$ of the other
component.}

\end{flushleft}
\end{table}

\begin{table}
\caption[]{Absolute surface flux of the different chromospheric indicators of GZ~Leo 
\label{tab:fluxf}}
\begin{flushleft}
\scriptsize
\begin{tabular}{lccccccccccc}
\noalign{\smallskip}
\hline \hline
\noalign{\smallskip}
 & &  \multicolumn{10}{c}{log$F$$_{\rm S}$ (P/S)}  \\
\cline{3-12}
\noalign{\smallskip}
 O & $\varphi$ &
 \multicolumn{2}{c}{Ca~{\sc ii}} & & & & & &
\multicolumn{3}{c}{Ca~{\sc ii} IRT}  \\
\cline{3-4}\cline{10-12}
\noalign{\smallskip}
 & &
 K   & H  & H$\epsilon$ & H$\delta$ & H$\gamma$ & H$\beta$ & H$\alpha$ &
$\lambda$8498 & $\lambda$8542 & $\lambda$8662
\tiny       
\\
\noalign{\smallskip}
\hline
\noalign{\smallskip}
M & 0.12 & - &  - & - & - & - & - & 6.52/6.54 & 6.37/6.19 & 6.23/6.21 & 6.3
1/6.31 \\
M & 0.77 & - &  - & - & - & - & - & 6.40/6.67 & 6.25/6.41 & 6.50/6.61 & 6.4
6/6.49 \\
M & 0.46 & - &  - & - & - & - & - & 6.52/6.68 &  $^{c}$/6.58 &  6.84$^{a}$/$^{c}$ & 6.62$^{a}$/$^{c}$ \\
M & 0.11 & - &  - & - & - & - & - & 6.52/6.51 & - & - & -  \\
M & 0.77 & - &  - & - & - & - & - & 6.48/6.59 & 6.00/6.27 & 6.41/6.48 & 6.3
4/6.35 \\
M & 0.42 & - &  - & - & - & - & - & 6.55/6.57 & 6.21/6.29 & 6.35/6.29 & 6.3
2/6.42 \\
M & 0.08 & - &  - & - & - & - & - & 6.59/6.66 & 6.04/6.15 & 6.43/6.37 & $^{c}$/$^{c}$ \\
M & 0.73 & - &  - & - & - & - & - & 6.58/6.73 & 6.00/6.34 & 6.40/6.46 & 6.2
7/6.25 \\
M & 0.39 & - &  - & - & - & - & - & 6.70/6.64 & 6.25/6.19 & 6.46/6.47 & 6.0
9/6.27 \\
M & 0.04 & - &  - & - & - & - & - & 6.67/6.51 & 6.15/6.40 & 6.32/6.47 & 6.2
3/6.50 \\
F & 0.68 & 7.11/7.09 &  7.01/7.05$^{e}$ &  $^{d}$/6.78 & 6.04/6.20$^{b}$ & $^{c}$/5.98$^{2}$ & 6.15/5.98$^{b}$ & 6.79/6.74 & 6.41/6.43 & 6.41/6.37 & 6.46/6.43 \\
F & 0.36 & 6.97/7.01 & 6.99$^{e}$/6.92 & 6.32/$^{d}$ & 5.78/6.04$^{b}$ & 5.98/6.08$^{b}$ & 6.03/6.25$^{b}$ & 6.57/6.78 & 6.48/6.49 & 6.41/6.45 & 6.42/6.51 \\
F & 0.02 & 7.27$^{a}$ & 7.19$^{a}$ & 6.92$^{a}$ & 6.08$^{a,}$$^{b}$ & 6.08$^{a,}$$^{b}$ & 6.40$^{a,}$$^{b}$ & 6.93$^{a}$ & 6.67$^{a}$ & 6.59$^{a}$ & 6.78$^{a}$ \\
F & 0.70 & 7.05/7.02 & 7.05/6.86 & 6.39/6.63 & 5.93/5.98$^{b}$ & 6.16/6.03$^{b}$ & 6.33/6.28$^{b}$ & 6.75/6.69 & 6.45/6.46 & 6.41/6.40 & 6.50/6.47 \\
F & 0.71 & 6.97/6.88 & 6.93/6.92 & 6.69/6.73 & 5.98/5.98 & 5.98/6.03 & 6.19/6.19 & 6.76/6.72 & 6.45/6.46 & 6.43/6.40 & 6.49/6.46 \\
\noalign{\smallskip}
\hline
\noalign{\smallskip}
\end{tabular}

{
 Notes as in Table~\ref{tab:ewf}}.
\\
\end{flushleft}
\end{table}

\end{document}